\documentclass[%
 reprint,
 superscriptaddress,
 amsmath,amssymb,
 aps,
]{revtex4-2}

\usepackage{braket}
\usepackage{comment}
\usepackage{graphicx}
\usepackage{dcolumn}
\usepackage{bm}
\usepackage{mathrsfs}
\usepackage{hyperref}
\usepackage{notes2bib}

\newcommand{\Figref}[1]{Fig.~\ref{fig: #1}}

\newcommand{\Secref}[1]{Sec.~\ref{sec: #1}}
\newcommand{\Appref}[1]{Appendix~\ref{app: #1}}

\newcommand{\mbf}{\mathbf}
\newcommand{\mcl}{\mathcal}

\newcommand{\br}[1]{\left( #1 \right)}
\newcommand{\Br}[1]{\left\{ #1 \right\}}
\newcommand{\BR}[1]{\left[ #1 \right]}
\newcommand{\ev}[1]{\left\langle #1 \right\rangle}
\newcommand{\av}[1]{\left| #1 \right|}

\newcommand{\uar}{\uparrow}
\newcommand{\dar}{\downarrow}

\newcommand{\im}{i}
\newcommand{\ex}{e}
\newcommand{\alp}{\alpha}

\newcommand{\eps}{\epsilon}

\newcommand{\lam}{\lambda}

\newcommand{\veps}{\varepsilon}

\newcommand{\ham}{\mcl{H}}

\newcommand{\pri}{\prime}

\newcommand{\vS}{\bm{S}}

\newcommand{\iB}{\mathrm{B}}
\newcommand{\iE}{\mathrm{E}}
\newcommand{\iK}{\mathrm{K}}
\newcommand{\iS}{\mathrm{S}}
\newcommand{\Hc}{\mathrm{H.\,c.}}
\newcommand{\gs}{\mathrm{gs}}
\newcommand{\pf}{\mathcal{F}}
\newcommand{\ot}{\mathcal{T}}
\newcommand{\pd}{\mathcal{P}}

\newcommand{\smz}{\widetilde{\gamma}}
\newcommand{\wmz}{\widetilde{\gamma}_\bullet}

\begin{document}


\title{Coexistence of Strong and Weak Majorana Zero Modes \\in Anisotropic XY Spin Chain with Second-Neighbor Interaction}

\author{Kazuhiro Wada}
\affiliation{Department of Applied Physics, Tokyo University of Science, Tokyo 125-8585, Japan}%
\author{Takanori Sugimoto}
\email{sugimoto.takanori@rs.tus.ac.jp}
\affiliation{Department of Applied Physics, Tokyo University of Science, Tokyo 125-8585, Japan}%
\affiliation{Advanced Science Research Center, Japan Atomic Energy Agency, Tokai, Ibaraki 319-1195, Japan}
\author{Takami Tohyama}
\affiliation{Department of Applied Physics, Tokyo University of Science, Tokyo 125-8585, Japan}%

\begin{abstract}
We theoretically investigate Majorana zero modes emerging in an anisotropic XY spin chain with second neighbor interactions.
The spin chain is mathematically equivalent to the Kitaev chain composing of spinless fermions if only nearest-neighbor interactions are considered.
Recent studies on the Kitaev chain with long-ranged interactions, have presented coexistence of several Majorana zero modes.
Investigating the topological phase diagram of the anisotropic XY spin chain with second neighbor interactions, we find coexistence of several Majorana zero modes similar to the Kitaev chain.
However, we confirm that one of zero modes is restricted into a Hilbert subspace.
The mode is regarded as a so-called weak zero mode that should exhibit thermodynamical properties different from a (strong) zero mode appearing in the whole Hilbert space.
Since the quantum statistics of spin is the same as a hard-core boson, we expect an experimental realization of the weak mode in not only quantum spin materials but also optical lattices for cold atom systems.
\end{abstract}

\pacs{Valid PACS appear here}
\maketitle



\section{Introduction}
Majorana zero mode (ZM) is a zero-energy excitation defined as a Majorana fermion, whose anti-particle corresponds to itself~\cite{Wilczek2009,Alicea2012,Elliott2015}.
In condensed matter physics, the Majorana ZM emerging in topological superconductors~\cite{Read2000,Kitaev2001} has attracted much attention, because it not only gives fundamental and valuable information of quantum properties originating from non-abelian statistics~\cite{Ivanov2001}, but also has a capability to carry out topologically protected quantum computing~\cite{Sau2010,Wieckowski2020} and Majorana qubits~\cite{Goldstein2011,Cheng2012,Budich2012,Rainis2012,Schmidt2012,Mazza2013}.
For these purposes, realization and control of the Majorana ZMs are urgent issues in current technologies of nano fabrication from both theoretical and experimental points of view~\cite{Shivamoggi2010,Oreg2010,Lutchyn2010,Alicea2010,Cook2012,Deng2012,Mourik2012,Das2012,Rokhinson2012,Sau2012,Finck2013,Churchill2013,Nadj-Perge2014,Okamoto2014,Higginbotham2015,Malard2016,Sato2016,Molignini2017,Molignini2018,Jack2019,Pan2019,Frolov2020,Molignini2020}, and moreover these studies bring numerous discussions on the effects of dimerization, quasiperiodicity, disorder and interaction~\cite{Motrunich2001,Brouwer2011a,Brouwer2011,Akhmerov2011,Lang2012,Tezuka2012,Lobos2012,Niu2012,Lobos2012,Degottardi2013,Wakatsuki2014,Adagideli2014,Hegde2016,Gergs2016,Zhu2016,Gergs2016,Herviou2016,Wang2017,Ghadimi2017,Perez2017,Nava2017,McGinley2017,Shapourian2017,Miao2017,Ezawa2017,Dey2017,Hung2017,Wang2018,Lieu2018,Thakurathi2018,Monthus2018,Wieckowski2018,Li2018,Wouters2018,Thakurathi2018,Kells2018,Miao2018,Habibi2018,Levy2019,Hua2019,Wieckowski2019,Griffith2020,Kobiaka2020,Mahyaeh2020,Aksenov2020,Wang2020,Yu2020,Fendley2016,Sugimoto2017,Chitov2018,Zheng2019,SAHA2019257,Chitov2019,Magnifico2019,Pandey2020,Sarkar2020,Yates2020,Yates2020a,Magnifico2020,Kumar2021}.

As a typical topological superconductor, the Kitaev chain is well investigated~\cite{Kitaev2001}.
In this model, a pairing potential between spinless fermions on neighboring sites plays a key role in topological superconductivity.
An alternative to the Kitaev chain is an anisotropic XY spin chain (AXYSC) with $S=1/2$ spins given by the Jordan--Wigner transformation of the Kitaev chain~\cite{Jordan1928,Fendley2016,Sugimoto2017,Chitov2018,Zheng2019,SAHA2019257,Chitov2019,Pandey2020,Yates2020,Yates2020a,Kumar2021,Sarkar2020}.
These models are mathematically equivalent if they have only nearest-neighbor interactions.

Recent studies on long-ranged interactions in the Kitaev chain~\cite{Niu2012,Jafari2016,Habibi2018a,Habibi2018,Lieu2018} have reported coexistence of several Majorana ZMs~\cite{Jafari2016,Lieu2018,Habibi2018}.
The increase of Majorana ZMs is in general important for improving efficiency of quantum computing.
From experimental point of view, long-ranged spin interactions, e.g., super-exchange couplings, the RKKY interactions~\cite{Ruderman1954,Kasuya1956,Yosida1957}, and dipole-dipole interactions, are more easily introduced as compared with long-ranged superconducting pair potential of spinless fermions appearing in the Kitaev chain.
Therefore, in this paper, we investigate the effects of long-ranged interactions in the spin chain, which differ from the long-ranged interactions in the Kitaev chain.

Besides, in connection with non-Abelian statistics, another type of Majorana ZM has also attracted much attention in recent years~\cite{Fendley2012,Jermyn2014,Alicea2016,Fendley2016,Wouters2018}.
The Majorana ZM, so-called {\it weak} Majorana ZM, is restricted into a Hilbert subspace including low-lying eigenstates, while the conventional Majorana ZM called {\it strong} Majorana ZM emerges in the whole Hilbert space.
The weak Majorana ZM, which is basically discussed in non-Abelian parafermion systems, behaves in the same manner with the strong Majorana ZM at low temperature.
Since the parafermions satisfy an intermediate commutation relation between fermions and hard-core bosons, clarifying behaviors of Majorana ZMs in a spin chain obeying the hard-core bosonic commutation relation is an important study from the quantum statistical point of view.

In contrast to the Kitaev chain, the effects of long-ranged interactions in the spin chain cannot be easily examined due to many-body interactions originating from the Jordan--Wigner string.
To overcome this difficulty, we combine numerical calculation based on variational matrix-product state (VMPS) method~\cite{Schollwock2011} with an analytical calculation in the Ising limit.
Note that the concept of the VMPS is based on the density-matrix renormalization group (DMRG) method proposed by S. White~\cite{White1992,White1993}, and thus, the VMPS is mathematically equivalent to the DMRG. 
In the numerical algorithm, the VMPS, however, has some differences from the DMRG, e.g., the singular-value decomposition of coefficient rectangular matrix of the wave function instead of diagonalization of the reduced density matrix, and matrix-product operator (MPO) representation of the Hamiltonian as a more sophisticated form of renormalized operators (see \Appref{MPO representation} for the AXYSC model).
While the VMPS method gives only a few low-lying eigenstates, the analyses of the Ising limit provides all eigenstates because of classical level for special parameters.
Combining both the calculations, we can discuss the overall structure in the full Hilbert space to clarify the Majorana ZM.

In this paper, we start with the correspondence between the Kitaev and Ising chains.
The Majorana ZM is derived in the Ising chain, and we discuss its features about boundary conditions.
In addition, we analytically show a weak Majorana ZM emerging in the Ising chain with second neighbor interactions.
This weak Majorana ZM originating from extension of unit cell in the ground states is induced by the second-neighbor interactions.
Next, to investigate the effects of quantum fluctuations, we present the VMPS study on low-lying eigenstates in the AXYSC with second neighbor interactions.
Based on detailed examinations of degeneracy in the low-lying states, we determine the topological phase diagram.
To confirm robustness of the weak Majorana ZM against quantum fluctuations, we numerically clarify energy spectra in several parameter points.
In these calculations, we find a non-topological phase, where there is no Majorana ZMs but a weak {\it trivial} ZM associated with extension of unit cell as well as the doubly-degenerate Majumdar--Ghosh ground states~\cite{Majumdar1969,Majumdar1970,Hikihara2001}.
Finally, we summarize these results and mention possible experiments to find the weak Majorana ZM.

The contents of this paper are as follows.
In Sec.~II, we introduce the model Hamiltonians of the AXYSC and the Kitaev chain together with the Jordan--Wigner transformation.
In Sec.~III, the Majorana ZMs in the Ising chain are exactly derived as an example of the AXYSC.
Two types of Majorana ZMs, i.e., strong and weak Majorana ZMs, are demonstrated in the Ising chain.
After explaining the VMPS method, we present topological phase diagram obtained by the VMPS method in Sec.~V.
As another VMPS result, we also show degeneracy in ground states in Sec.~VI.
Sections VII and VIII are used for discussion and summary, respectively.


\section{Model}
\label{sec: model}
We consider the AXYSC with a second neighbor interaction given by
\begin{equation}
\label{eq: Ham}
\ham_\iS = \sum_{n=1,2} \BR{\ham_n^{\iB} + \eta\ham_n^{\iE}},
\end{equation}
with
\begin{align}
\label{eq: Ham-b}
\ham_n^{\iB} &= \sum_{j=1}^{N-n} \BR{\br{ J_n + \lambda_n } S_j^x S_{j+n}^x + \br{ J_n - \lambda_n } S_j^y S_{j+n}^y},\\
\label{eq: Ham-e}
\ham_n^{\iE} &= \sum_{j=1}^{n} \BR{\br{ J_n + \lambda_n} S_{N-n+j}^x S_{j}^x + \br{ J_n - \lambda_n } S_{N-n+j}^y S_{j}^y},
\end{align}
where $\ham_n^{\iB}$ ($\ham_n^{\iE}$) is the bulk (edge) Hamiltonian of $n$-th neighbor interaction for $n=1,2$ (see \Appref{MPO representation} for the MPO representation). The $x$ ($y$) component of spin-$1/2$ operator on $j$-th site is denoted by $S_j^x$ ($S_j^y$). The exchange energy and anisotropy of $n$-th neighbor coupling are defined as $J_n$ and $\lambda_n$, respectively [see \Figref{model}]. To control the boundary condition, we use the parameter $\eta$.

\begin{figure}[h]
	\centering
	\includegraphics[width=0.95\linewidth]{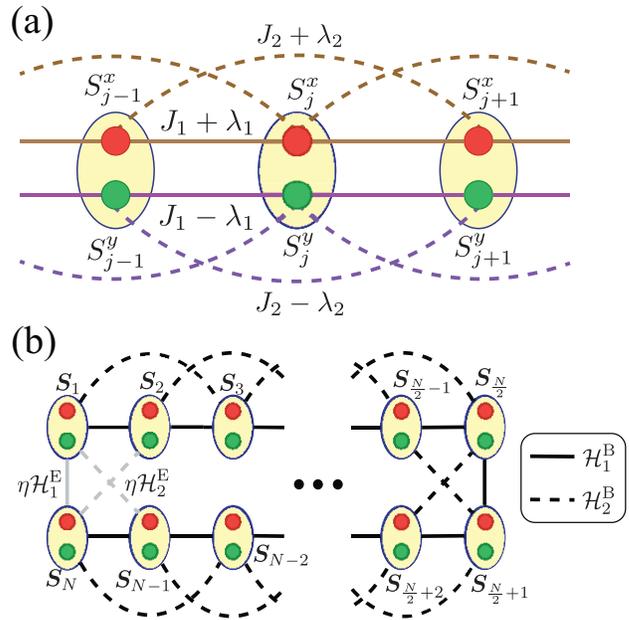}
	\caption{Schematics of the AXYSC. Red (green) balls represent $x$ ($y$) components of spin-$1/2$ operator. The spin operator except for $z$ component is denoted by a yellow oval. Solid and dashed lines are the first and second neighbor interactions, respectively. (a) The anisotropic interactions of $x$ ($y$) components are colored brown (purple). (b) The bulk (edge) Hamiltonian is colored black (gray). Without the edge Hamiltonian ($\eta=0$), the system corresponds to the open boundary condition (PBC) , while the periodic boundary condition (PBC) is chosen with $\eta=1$.}
	\label{fig: model}
\end{figure}

The Majorana ZM emerges on the edges in a topological phase of one-dimensional $p$-wave superconductor (the BdG class)~\cite{Schnyder2008}, e.g., the Kitaev chain, with the open boundary condition (OBC). The Kitaev chain of $n$-th neighbor interaction is given by
\begin{equation}
\label{eq: Kitaev}
\ham_n^{\iK} = \sum_{j=1}^{N-n} \br{t_n c^\dagger_j c_{j+n} + \Delta_n c^\dagger_j c^\dagger_{j+n} + \Hc},
\end{equation}
where $c_j^\dag$ ($c_j$) denotes the creation (annihilation) operator of spinless fermion on $j$-th site.
The Kitaev chain with only the first neighbor interaction $\ham_1^{\iK}$ is mathematically equivalent to the AXYSC $\ham_1^{\iB}$ with $2t_1=J_1$ and $2\Delta_1=\lambda_1$ through the Jordan--Wigner transformation~\cite{Jordan1928},
\begin{equation}
S^+_j = c_j^\dag \prod_{k=1}^{j-1} \ex^{-\im\pi n_k} ,\quad S^-_j = c_j \prod_{k=1}^{j-1} \ex^{\im\pi n_k},
\end{equation}
where $n_j=c_j^\dag c_j$ represents the number operator of spinless fermion on $j$-th site.
The topological bulk state in the Kitaev chain has a finite energy gap, so that the Majorana ZM is also expected in the AXYSC model with small $J_2$ and $\lambda_2$ as compared with $J_1$ and $\lambda_1$ under OBC ($\eta=0$).
Furthermore, preceding works have pointed out that a second neighbor interaction $\ham_2^{\iK}$ in the Kitaev model induces multiple Majorana ZMs~\cite{Jafari2016,Lieu2018,Habibi2018}.
However, the second neighbor interaction $\ham_2^{\iB}$ in the AXYSC model discords with $\ham_2^{\iK}$ in the Kitaev model, because the Jordan--Wigner phase of a middle site, so-called Jordan--Wigner string, does not cancel out after the transformation, that is,
\begin{equation}
\ham_2^{\iB} = \frac{1}{2}\sum_{j=1}^{N-2} \br{J_2 c_j^\dag \ex^{\im \pi n_{j+1}} c_{j+2} + \lambda_2 c_j^\dag \ex^{\im \pi n_{j+1}} c_{j+2}^\dag + \Hc}.
\end{equation}
Thus, it is an interesting problem to clarify whether $\ham_2^{\iB}$ in the AXYSC model also induces multiple Majorana ZMs as well as $\ham_2^{\iK}$ in the Kitaev chain.


\section{Majorana ZMs in Ising chain}
\label{sec: Majorana ZM}
We first demonstrate the Majorana ZM appearing in the Ising chain as the classical limit of the AXYSC.
The Majorana ZM is a zero-energy quasiparticle excitation which obeys the Majorana condition, i.e., its Hermitian conjugate is the same as itself.
Two types of Majorana operators $\gamma_j^\tau$ ($\tau=a,b$) satisfying the Majorana condition $\gamma_j^\tau = \br{\gamma_j^\tau}^\dag$ are introduced by one fermion:
\begin{equation}
\label{eq: Majorana operators}
\gamma_j^a = \frac{1}{\sqrt{2}}\br{c_j^\dag + c_j}, \quad \gamma_j^b = \frac{\im}{\sqrt{2}} \br{c_j^\dag - c_j}.
\end{equation}
These operators obey the fermionic anticommutation relation $\{ \gamma_j^\tau , \gamma_{j^\prime}^{\tau^\prime} \} = \delta_{j,j^\prime}\delta_{\tau,\tau^\prime}$.
Moreover, the Majorana operator can be regarded as an operator which changes the fermion-number parity given by
\begin{equation}
\label{eq: fermionic parity}
\pf = \ex^{\im \pi \sum_j n_j} = \prod_j 2\im \gamma_j^b \gamma_j^a.
\end{equation}
In fact, the Majorana operators anticommute with the fermionic parity $\Br{ \gamma_j^\tau , \pf } = 0$, and the fermionic parity is conserved in the AXYSC, $[\pf,\ham_\iS]=0$.
Thus, with considering a simultaneous eigenstate of the Hamiltonian and the fermionic parity $\ket{\varepsilon,\chi}$ with eigen-energy $\varepsilon$ and eigenvalue of the fermionic parity $\chi=\pm$, the state obtained by the Majorana operator acting on an eigenstate has an opposite parity to the eigenstate:
\begin{equation}
  \pf\br{\gamma_j^\tau\ket{\varepsilon,\chi}}=-\chi \br{\gamma_j^\tau\ket{\varepsilon,\chi}}.
\end{equation}
Furthermore, if there is a Majorana operator $\smz$ commuting with the Hamiltonian $[\smz,\ham_\iS]=0$, a pair of eigenstates with different fermionic parity exists at every energy level, $\ket{\varepsilon,\chi}$ and $\ket{\varepsilon,-\chi}\equiv \sqrt{2}\smz\ket{\varepsilon,\chi}$.
This Majorana operator $\smz$ is called (strong) Majorana ZM.
Besides the commutation relation $[\smz,\ham_\iS]=0$, recent theoretical study has predicted the Majorana ZM satisfying a weaker condition $\pd[\wmz,\ham_\iS]\pd=0$ with a projection to a certain Hilbert subspace $\pd$~\cite{Fendley2012,Jermyn2014,Alicea2016,Fendley2016,Wouters2018}.
In this case, only eigenstates belonging to the subspace are doubly degenerate.
The Majorana ZM $\wmz$ with such a weaker condition $\pd[\wmz,\ham_\iS]\pd=0$ is called {\it weak} Majorana ZM, as compaired with the strong Majorana ZM $\smz$ satisfying $[\smz,\ham_\iS]=0$.

We can easily find the strong Majorana ZM in the Ising chain without the second neighbor interaction, corresponding to a condition $\lambda_1=\pm J_1$ in the AXYSC.
For instance, the Majorana representation of the Ising chain is given by,
\begin{equation}
\ham_1^{\iB}|_{\lambda_1=J_1} = -J_1 \sum_{j=1}^{N-1} \im \gamma_j^b \gamma_{j+1}^a.
\end{equation}
Two Majorana operators $\gamma_1^a$ and $\gamma_N^b$ on the edges apparently commute with the Hamiltonian, indicating two strong Majorana ZMs $\smz^a=\gamma_1^a$ and $\smz^b=\gamma_N^b$ existing in the Ising chain.
These Majorana ZMs $\smz^a$ and $\smz^b$ are essentially the same Majorana ZM.
The reason is as follows.
The Hermite operator $2\im \smz^a \smz^b$, which has two eigenvalues $\chi_e=\pm 1$, commutes with both the Hamiltonian $\ham_1^{\iB}|_{\lambda_1=J_1}$ and the fermionic parity $\pf$, so that there are simultaneous eigenstates $\ket{\varepsilon,\chi,\chi_e}$.
With applying $\smz^a $ on the eigen-equation of $2\im \smz^a \smz^b$, we obtain the following relation.
\begin{equation}
  \smz^b\ket{\varepsilon,\chi,\chi_e}=-\im\chi_e\smz^a\ket{\varepsilon,\chi,\chi_e}.
\end{equation}
These two states are the same without the phase factor $-\im\chi_e$, and thus, the Majorana ZMs $\smz^a$ and $\smz^b$ are essentially equivalent (see \Appref{Majorana-pair} for more information of these Majorana ZMs).
Interestingly, these strong Majorana ZMs survive with periodic boundary condition (PBC) in the Ising chain, that is, the $\smz^a$ and $\smz^b$ also commute with the edge Hamiltonian,
\begin{equation}
\ham_1^{\iE}|_{\lambda_1=J_1} = -\im J_1 \gamma_N^b \gamma_1^a\pf.
\end{equation}
This startling feature originates from the operator $\pf$ as a remnant of the Jordan--Wigner transformation, and thus, this feature is inherent in the spin chain.
In a general case with $J_1 \neq \lambda_1$, a strong Majorana ZM is obtained as a superposition of Majorana operators.
In the Kitaev chain with OBC, this mode is slightly expanded from edges but exponetially decreasing into the bulk.
Thus, it is exactly obtained only in the thermodynamic limit.

Next, let us consider the effects of the second neighbor interaction in the Ising chain with PBC ($\lambda_1=J_1$ and $\lambda_2=J_2$).
In the following, the system size $N$ is set to satisfy $N=0\ (\mathrm{mod.}\ 4)$, and the interactions are restricted into antiferromagnetic $J_1>0$ and $J_2>0$ for simplicity.
The gound-state phase undergoes the first-order phase transition at $J_1^x=2J_2^x$, where the degeneracy of ground states changes from two-fold ($J_1^x>2J_2^x$) to four-fold ($J_1^x<2J_2^x$).
In the latter region $J_1^x<2J_2^x$, the four-fold ground states are given by
\begin{align}
\label{eq: four-fold degenerate states}
\ket{\gs_1} &= \ket{\uar \uar \dar \dar \cdots}_x, \quad \ket{\gs_2} = \ket{\dar \dar \uar \uar \cdots}_x, \\
\ket{\gs_3} &= \ket{\uar \dar \dar \uar \cdots}_x, \quad \ket{\gs_4} = \ket{\dar \uar \uar \dar \cdots}_x,
\end{align}
where $\ket{\uar \dar \uar \dar \cdots}_x$ is defined as a direct product, $\ket{\uar}_1^x\ket{\dar}_2^x\ket{\uar}_3^x\ket{\dar}_4^x \cdots$ with the eigenstates of $S_j^x$, $\ket{\uar}_j^x$ and $\ket{\dar}_j^x$.
To diagonalize the fermionic parity in these states, we adopt bonding and antibonding ground states as linear combinations of these states:
\begin{equation}
  \ket{\gs_{1,2}^{\pm}} = \frac{1}{\sqrt{2}} \br{\ket{\gs_1} \pm \ket{\gs_2}}, \quad \ket{\gs_{3,4}^{\pm}} = \frac{1}{\sqrt{2}} \br{\ket{\gs_3} \pm \ket{\gs_4}}.
\end{equation}
With the relations $\ket{\dar}_j^x=-\ex^{\im\pi n_j}\ket{\uar}_j^x$ and $\ket{\uar}_j^x=-\ex^{\im\pi n_j}\ket{\dar}_j^x$, where $\ex^{\im\pi n_j}$ corresponds to a spin-flip operator of $x$ component, eigen-equations of the fermionic parity are obtained by
\begin{equation}
\pf\ket{\gs_{1,2}^{\pm}}=\pm  \ket{\gs_{1,2}^{\pm}}, \quad \pf \ket{\gs_{3,4}^{\pm}}=\pm \ket{\gs_{3,4}^{\pm}}.
\end{equation}
Thus, there is a Majorana ZM between the bonding and antibonding ground states.
This chracteristic is common to all energy levels.
In fact, every state can be written by a direct product of local eigenstate $\ket{\uar}_j^x$ or $\ket{\dar}_j^x$, e.g., $\ket{\psi}=\ket{\uar \uar \dar \uar \cdots}_x$.
The direct product state $\ket{\psi}$ have a spin-flip pair $\pf\ket{\psi}=\ket{\dar \dar \uar \dar \cdots}_x$, which has the same energy as $\ket{\psi}$.
The bonding and antibonding states $\ket{\psi_\pm}=\br{\ket{\psi}\pm\pf\ket{\psi}}/\sqrt{2}$ have diferrent fermionic parity.
Therefore, this Majorana ZM between bonding and antibonding states at each energy level corresponds to a strong Majorana ZM, $\smz^a=\sum_\psi \br{\ket{\psi_+}\bra{\psi_-}+\ket{\psi_-}\bra{\psi_+}}/\sqrt{2}$, e.g., $\smz^a=\sqrt{2}S_1^x$.

On the other hand, there is another Majorana ZM in the ground states, i.e., a Majorana ZM between $\ket{\gs_{1,2}^{+}}$ and $\ket{\gs_{3,4}^{-}}$ ($\ket{\gs_{1,2}^{-}}$ and $\ket{\gs_{3,4}^{+}}$).
This mode is constructed by the spin-flip operator of odd sites $\pf_o$ and the parity-flip operator $S_1^x$, i.e., $\wmz^a=\sqrt{2}\im S_1^x\pf_o$.
This operator also satisfies the Majorana condition $\{\smz^a, \wmz^a\}=0$, $2(\wmz^a)^2=1$, and $\wmz^a= (\wmz^a)^\dag$.
However, for instance, with respect to the highest-energy (saturated) states $\ket{\uar \uar \uar \cdots}_x$ and $\ket{\dar \dar \dar \cdots}_x$, which are only doubly degenerate, $\wmz^a$ corresponds to a Majorana mode in a different energy sector.
The difference of degeneracy originates from $\pd[\wmz^a,\ham_\iS]\pd=0$ with a projection $\pd$ to a certain Hilbert subspace including the ground states but except for the saturated states.
Therefore, this mode is regarded as a weak Majorana ZM.

It is also worthly noted that this weak Majorana ZM is regarded as a product of the strong Majorana ZM $\smz^a$, the fermionic parity $\pf$, and a weak (not Majorana) ZM, that we call a weak {\it trivial} ZM in this paper, between $\ket{\gs_{1,2}^{+}}$ and $\ket{\gs_{3,4}^{+}}$ ($\ket{\gs_{1,2}^{-}}$ and $\ket{\gs_{3,4}^{-}}$).
This weak trivial ZM corresponds to a translation operator $\ot$ defined by
\begin{equation}
  \ket{\gs_1} = \ot \ket{\gs_3}, \quad \ket{\gs_2} = \ot \ket{\gs_4},
\end{equation}
In the ground states, the translation operator $\ot$ is equivalent to the spin-flip operator of even sites $\pf_e$, i.e., $\pd \ot \pd =\pd \pf_e \pd$.
Thus, we can obtain the relation $\wmz^a=\im \smz^a \pf \ot$ in the ground states.

More generally, if there is a trivial ZM $\zeta$ defined by any parity-flip operator ($\zeta^2=1$), commuting with another (intrinsic) Majorana ZM $\smz$, the fermionic parity $\pf$, and the Hamiltonian $\ham$, i.e., $[\zeta,\smz]=[\zeta,\pf]=[\zeta,\ham]=0$, a product of them, $\smz^\pri=\im \smz\pf \zeta$, satisfies the Majorana ZM conditions independent from the $\smz$, namely $\{\smz^\pri,\smz\}=\{\smz^\pri,\pf\}=[\smz^\pri,\ham]=0$, $2(\smz^\pri)^2=1$, and $\smz^\pri= (\smz^\pri)^\dag$.
To avoid misunderstanding, we again insist that this Majorana ZM $\smz^\pri$ is diffrent from the pair to the Majorana ZM $\smz$ (see \Appref{Majorana-pair2} for more information).
Moreover, it is intriguing that this reconstructed Majorana ZM $\smz^\pri$ can exist only with the intrinsic Majorana ZM $\smz$, otherwise there is just a trivial ZM.
In the Ising chain with large enough second neighbor interactions, unit cell of the ground state is extended twice as large as the N\'eel state, resulting in generation of sublattice degree of freedom associated with the translation.
This degree of freedom corresponds to the weak trivial ZM ($\zeta$) in the ground states, so that we can find a weak Majorana ZM ($\smz^\pri$) if there is another strong Majorana ZM ($\smz$).


\section{Numerical Method}
\label{sec: method}
In general, the Majorana ZM in the AXYSC with second neighbor interactions is not easy to be clarified due to quantum fluctuation.
To clarify the quantum effects, we investigate low-lying states with the VMPS method.
In this study, we set bond dimension of the VMPS to be greater than 400 and confirmed that truncation error is approximately less than $10^{-6}$.
If there is a degeneracy in the ground states with different fermionic parities, we can confirm at least existence of a weak Majorana ZM.
Moreover, we can discuss the type of ZMs by comparison with the Ising chain.
The number and type of Majorana ZMs can help us to determine ground-state phase diagram.

In the AXYSC, the strong Majorana ZM is not localized in the edges, and thus, we can detect the Majorana ZM with both OBC and PBC by checking the degeneracy of the ground states. 
In addition, the bulk state is common between OBC and PBC in the thermodynamical limit, so that the topological phase diagrams with OBC and PBC should be the same.
Thus, to determine the topological phase diagram, we choose the OBC to reduce the numerical costs of the VMPS method. 
On the other hand, as explained in Sec.~III, the translational symmetry is crucial in finding the weak trivial ZM, irrespective of existence of the strong Majorana ZM.
Since the OBC breaks the translational symmetry, we apply the PBC when we discuss the existence of the weak trivial ZM. 
Note that even with OBC, we can find the phase transition related to the weak trivial ZM of PBC, because the first-excitation gap closes at the transition point.
If there is the strong Majorana ZM, the weak trivial ZM is regarded as an additional Majorana ZM (see \Appref{Majorana-pair2}).
Therefore, we can obtain a more detailed classification of topological phase diagram in light of the weak trivial ZM.

To obtain the ground-state energy in specified parity subspaces, we add an auxiliary field in the Hamiltonian defined by
\begin{equation}
  \ham_F=\mu\sum_\alpha \av{\pf^\alpha-\chi^\alpha},
\end{equation}
where $\pf^\alpha=\prod_{j=1}^N(-2S_j^\alpha)$ is the $\alpha=x,y,z$ component of parity operator, which is conserved in the Hamiltonian because of $[\pf^\alpha,\ham_S]=0$.
The fermionic parity $\pf$ in \Secref{Majorana ZM} corresponds to $z$ component $\pf^z$.
These parity operators have $Z_2$ quantities according to their eigenvalues, while there is a constraint $\pf^x\pf^y\pf^z=(-\im)^N$.
For simplicity, in this paper, we consider only $N=4n$ system size with a positive integer $n$, so that acceptable subspaces distinguished by eigenvalues of the parities, $\bm{\chi}_i=(\chi_i^x,\chi_i^y,\chi_i^z)$, are given by
\begin{align}
&\bm{\chi}_0=(+,+,+), \bm{\chi}_1=(+,-,-), \notag\\
&\bm{\chi}_2=(-,+,-), \bm{\chi}_3=(-,-,+),
\end{align}
where $\chi_i^\alpha$ denotes each eigenvalue of the parity operator $\pf^\alpha$~\footnote{In general, the Hilbert subspace in the anisotropic XYZ spin chain can be classified with only the fermionic parity $\pf=\pf^z$ and the spin-flip operator $\pf^x$, i.e., $\pf^y$ is not necessary. In this paper, however, we use a more symmetric form by introducing $\pf^y$. This representation is useful when we consider a corresponding model via a SU(2) rotation.}.
Therefore, with a fixed $\bm{\chi}$ for the target subspace, we can calculate the ground-state energy using the VMPS method for $\ham_S+\ham_F$ with a large enough positive potential $\mu>0$ (see \Appref{MPO representation} for the MPO representation).

Furthermore, to clarify the degeneracy in the same subspace, we also calculate the $\nu$-th lowest eigen-energy $\varepsilon_\nu$ after obtaining up to $(\nu-1)$-th low-lying eigenstates $\ket{l}$ ($l=0,1,2,\cdots,\nu-1 $).
In the VMPS method, we can obtain any excitation energy with shifting all low-lying states to upper levels.
To shift the low-lying states, we introduce additional term defined by
\begin{equation}
  \ham_L=\epsilon \sum_{l=0}^{\nu-1} \ket{l}\bra{l}.
\end{equation}
With applying an enegy shift larger than $\nu$-th excitation energy, $\epsilon>\varepsilon_{\nu}-\varepsilon_0$, the lowest-energy state corresponds to the $\nu$-th eigenstate, since the Hamiltoanin is deformed into $\ham_S+\ham_L=\sum_{l\geq \nu} \varepsilon_l \ket{l}\bra{l}+ \sum_{l< \nu} (\varepsilon_l+\epsilon) \ket{l}\bra{l}$.
Thus, if we separately set the shift energy to the eigen-energy, i.e., $\eps=-\veps_l$, the effect of $\ham_L$ corresponds to the projection into the Hilbert subspace perpendicular to the subspace composing of the low-lying eigen-vectors $\ket{l}$ ($l=0,1,\cdots,\nu-1$).


\section{Topological phase diagram}
\label{sec: phase diagram}

In this section, we present the topological phase diagram of the AXYSC, which is determined by the degeneracy and parities of ground states with OBC.
The degeneracy and parities are clarified by the VMPS method with the auxiliary field (see \Secref{method}).
Since the Majorana ZMs in the AXYSC emerge in the thermodynamic limit, we first confirm the size effects.

\begin{figure}[htpb]
  \centering
  \includegraphics[width=1.0\linewidth]{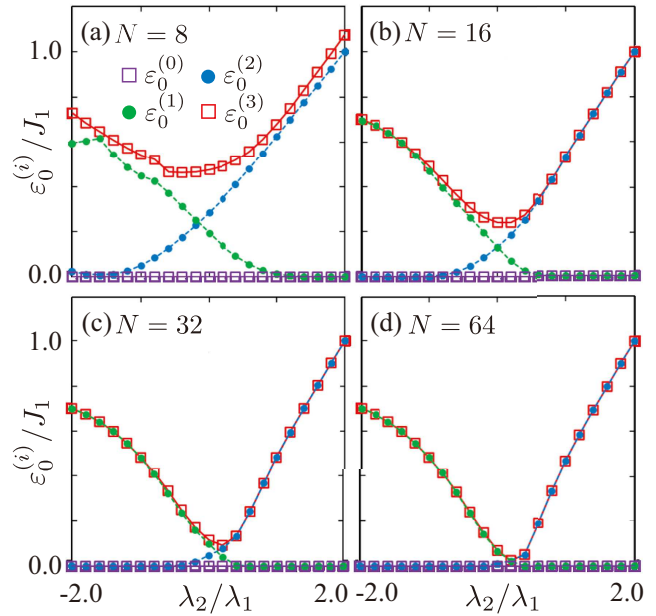}
  \caption{Size dependence of ground-state energies $\varepsilon_0^{(i)}$ in four subspaces with different parity $\bm{\chi}_i$ for $i=0,1,2,3$, corresponding to $(+,+,+), (+,-,-), (-,+,-), (-,-,+)$ with fixed $\lambda_1/J_1=1$ ($\theta=\pi/4$) and $J_2/J_1=2$ for (a) $N=8$, (b) $N=16$, (c) $N=32$, and (d) $N=64$ systems. The real ground-state energy $\varepsilon_0^{(0)}$ is set to be zero.}
  \label{fig: size-effect}
\end{figure}

Figure \ref{fig: size-effect} shows size dependence of the ground-state energies in four subspaces $\bm{\chi}_i$ ($i=0,1,2,3$), which are identified by the 3-component parities $\pf^\alpha$.
With increasing the system size $N$, i.e., from (a) to (d), the energies converge into two levels with two-fold degeneracy. In (d), the ground states in $\bm{\chi}_0$ and $\bm{\chi}_2$ ($\bm{\chi}_1$ and $\bm{\chi}_3$) are degenarate for $\lambda_2/\lambda_1\lesssim 0.2$, while the ground states in $\bm{\chi}_0$ and $\bm{\chi}_1$ ($\bm{\chi}_2$ and $\bm{\chi}_3$) are degenarate for $\lambda_2/\lambda_1\gtrsim 0.2$.

Next, to clarify the size dependence in more details, we show the ground-state energies as a function of inverse system size $1/N$, in which we can apparently check the thermodynamical limit.
In \Figref{SizeExtrapolation}, the size dependence of ground-state energies in subspaces is shown at certain parameter points, $(\theta,J_2/J_1,\lambda_2/\lambda_1)=(\pi/4,2.0,0.8)$, $(\pi/4,2.0,0.2)$, and $(7\pi/20,1.2,0.2)$.
In \Figref{SizeExtrapolation}(a), we can see the convergence of energies into two energy levels corresponding to the strong Majorana ZM.
On the other hand, we can see the convergence of energies into one energy level in \Figref{SizeExtrapolation}(b) (see also \Figref{size-effect}), indicating the phase boundary.
Moreover, we can find parameter points where no degeneracy exists at the ground state like \Figref{SizeExtrapolation}(c), implying no Majorana ZMs.
Based on these behaviors, we have classified topological phase diagram.

\begin{figure*}[htb]
	\centering
	\includegraphics[width=0.7\linewidth]{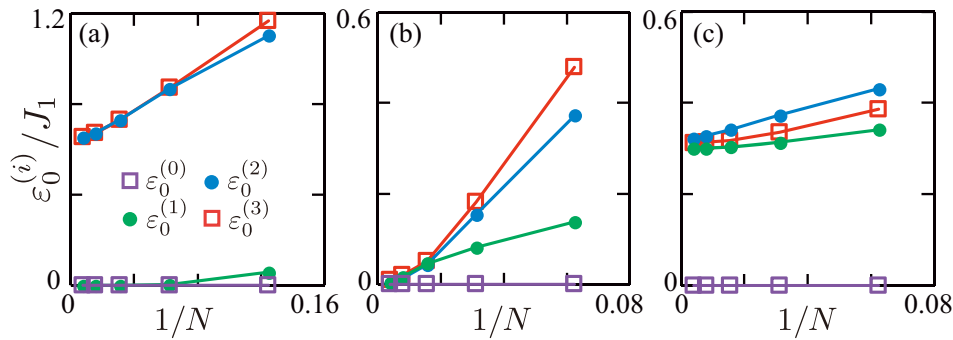}
	\caption{Size dependence of ground-state energies $\varepsilon_0^{(i)}$ in four subspaces $\bm{\chi_i}$. The parameter points are $(\theta,J_2/J_1,\lambda_2/\lambda_1)=$ (a) $(\pi/4,2.0,0.8)$, (b) $(\pi/4,2.0,0.2)$, and (c) $(7\pi/20,1.2,0.2)$.}

	\label{fig: SizeExtrapolation}
\end{figure*}

Since the parity component associated with degeneracy changes at the topological phase boundary, the four-fold degeneracy in the ground states helps us draw a topological ground-state phase diagram.
In \Figref{phase-diagram}, we show the phase diagram, where there are three parameters $\lambda_1/J_1$ represented by $\theta=\tan^{-1}(\lambda_1/J_1)$, $J_2/J_1$, and $\lambda_2/\lambda_1$.
We find three phases distinguished by the degeneracy and the parity components: two topological phases with different parities [(light and dark) green, and blue regions] and a non-topological phase without degeneracy between different parities (orange region).
In the (light and dark) green regions, the ground states are degenerate between $\bm{\chi}_0$ and $\bm{\chi}_1$ subspaces, while in the blue region, the ground-state degeneracy exists between $\bm{\chi}_0$ and $\bm{\chi}_2$ subspaces.
On the other hand, we do not find the ground-state degeneracy between different parity subspaces in the orange region, indicating that two-body interactions in second-neighbor terms break topological ground state and Majorana ZMs.

\begin{figure}[htpb]
	\centering
	\includegraphics[width=1.0\linewidth]{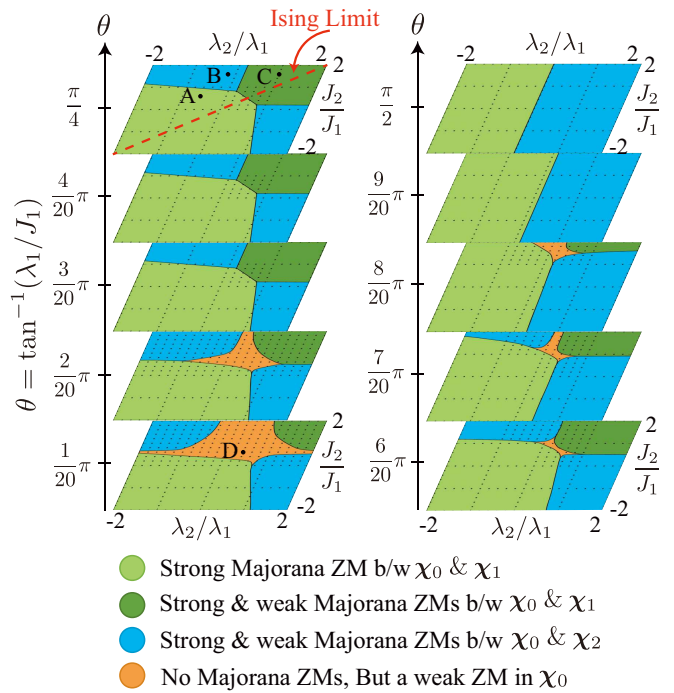}
	\caption{Topological phase diagram of the AXYSC with second-neighbor interaction. In the light and dark green regions, the ground state exhibits the Majorana ZM between $\bm{\chi}_0$ and $\bm{\chi}_1$ subspaces, whereas in the blue region, the Majorana ZM exists between $\bm{\chi}_0$ and $\bm{\chi}_2$ subspaces. Except for a common fermionic parity $\pf^z$, the green and blue topological phases are characterized by the fermionic parities $\pf^y$ and $\pf^x$, respectively. In the orange region, there is no degeneracy between different parity subspaces. The calculation is performed for an $N=64$ system with OBC. The parameter points A ($\theta=\pi/4, J_2/J_1=0.6$, $\lambda_2/\lambda_1=-0.6$), B ($\theta=\pi/4, J_2/J_1=1.6$, $\lambda_2/\lambda_1=-0.2$), C ($\theta=\pi/4, J_2/J_1=1.6$, $\lambda_2/\lambda_1=1.0$), and D ($\theta=\pi/20, J_2/J_1=0.6$, $\lambda_2/\lambda_1=0.4$) are used to demonstrate correlation functions (see \Figref{correlation}) and degeneracy at the ground states (see Sec.~VI). Red dashed line denotes the Ising chain with second-neighbor interactions, where $J_2/J_1=\lambda_2/\lambda_1$ at $\theta=\pi/4$ ($\lambda_1/J_1=1$).}
	\label{fig: phase-diagram}
\end{figure}

Next, we explain equivalence of separated blue regions and difference of two green regions.
On $\theta=\pi/4$, corresponding to $J_1=\lambda_1$, there is a mirror symmetry with respect to the diagonal line (Ising limit), because the Hamiltonian $\ham_2^\iB$ is invariant under permutation of $J_2$ and $\lambda_2$ with a distinct spin rotation around $x$ axis $S_j^y\to -S_j^y$ for $j=0,1 \ (\mathrm{mod.}\> 4)$.
Note that the spin rotation also keeps the parity operators $\pf^\alpha$.
Accordingly, the left-upper topological phase (a blue region) is identical to the right-lower topological phase (the other blue region) at least on $\theta=\pi/4$.
Since every blue region on each $\theta$ panel should be continuously connected to each other, we conclude that blue regions are the same phase.

However, recalling the phase transition in the Ising chain with the PBC, the light and dark green regions are not regarded as the same phase.
As explained in Sec.~III, the ground state in the Ising chain comes across the phase transition at $J_1^x=2J_2^x$, corresponding to the crossing point of the black line and the red dashed line on $\theta=\pi/4$ in \Figref{phase-diagram}, where degeneracy of the ground states changes from two-fold to four-fold with increasing $J_2=\lambda_2$.
In fact, we find that there is a sign of topological transition on the phase boundary between light and dark green regions, where all ground states in different four parity subspaces are degenerate (not shown) as for the phase boundaries between blue and green regions.

To confirm the phase transition accompanying the doubly-extended unit cell with spontaneous breaking of translational symmetry, we show the correlation functions of $\alp=x,y$ components of spins defined by
\begin{equation}
  C_\alp(r) = \ev{S_{N/2}^\alp S_{N/2+r}^\alp}.
\end{equation}
The correlation functions at the four parameter points A, B, C, and D in \Figref{phase-diagram}, representing the light green, blue, dark green, and orange regions, respectively.
By comparison between A and C in \Figref{correlation}, we can see that the periods of $C_x(r)$ are different.
In addition, we can also confirm that the $y$ component of the spin correlations $C_y(r)$ shows a long-range order in B, while in A and C, the $x$ component $C_x(r)$ is long-ranged.
On the other hand, in D, there is no long range order of spins, because both the correlation functions $|C_\alp(r)|$ exponentially decrease with increasing the length $r$.

\begin{figure}[h]
	\centering
	\includegraphics[width=\linewidth]{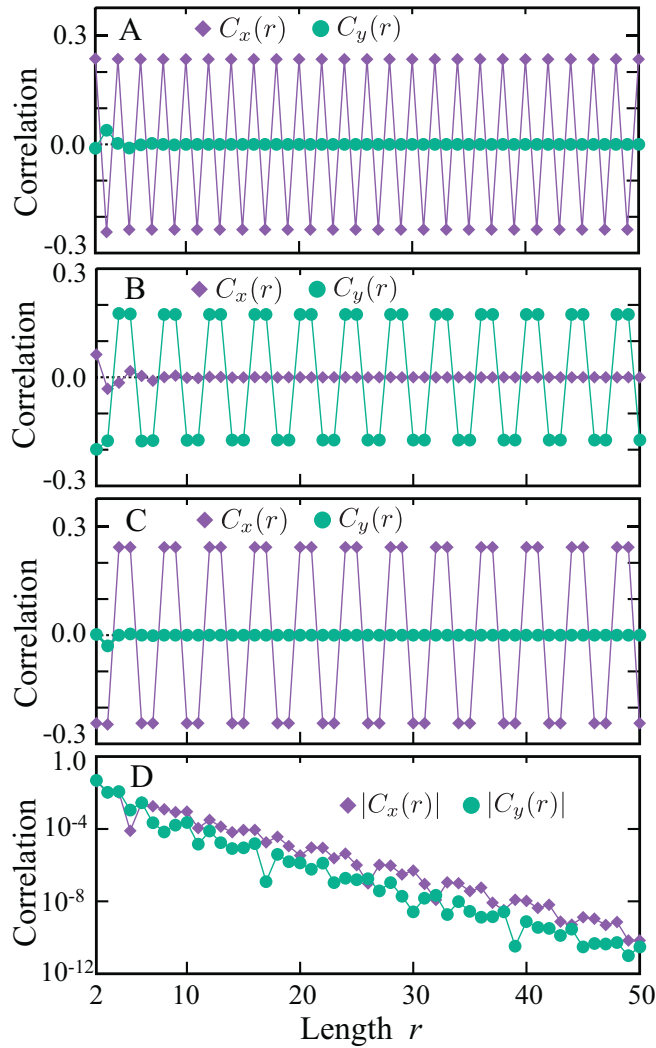}
	\caption{Correlation functions $C_\alp(r)$ for $\alp=x,y$ in the ground state at the four parameter points A, B, C, and D in \Figref{phase-diagram}. We calculate the ground state without any auxiliary fields $\ham_F$ and $\ham_L$, i.e., $\mu=\eps=0$, in an $N=256$ system with OBC.}
	\label{fig: correlation}
\end{figure}

Although the weak Majorana ZM is also expected off the diagonal line (Ising limit) in the dark green region, it is difficult to prove it analytically.
In the next section, we demonstrate the ground-state degeneracy at four points A, B, C, and D, representing the light green, blue, dark green, and orange regions, respectively.


\section{Ground-state degeneracy and weak Majorana ZM}
\label{sec: XY model}

To confirm the degeneracy associated with the weak Majorana ZM in topological phases, we numerically obtain lowest energies with PBC at three points A, B, and C at $\theta=\pi/4$ ($J_1=\lambda_1$) in \Figref{phase-diagram} by means of the VMPS method.

\begin{figure}[h]
	\centering
	\includegraphics[width=\linewidth]{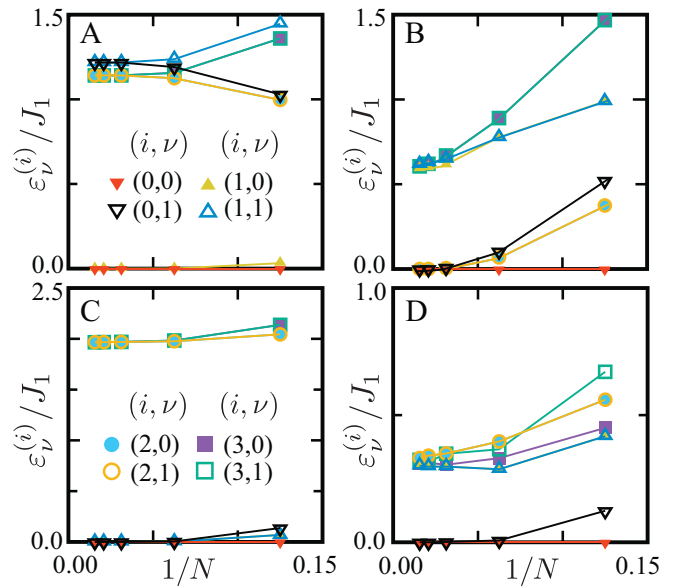}
	\caption{Size dependence of the ground-state ($\nu=0$) and first-excitation ($\nu=1$) energies in four subspaces $\bm{\chi}_i$ $(i=0,1,2,3)$, by means of the VMPS method in the $N=8,16,32,48,64$ AXYSCs with the PBC. Four panels correspond to the points A, B, C, and D in \Figref{phase-diagram}. In each panel, closed (open) symbols denote the ground-state (first-excitation) energies with four different shapes, upward triangle, downward triangle, circle, and square corresponding to the subspace indeces $i=0,1,2$, and $3$, respectively. The energy origin is set to the minimal energy of the ground-state energies.}
	\label{fig: size-dependence}
\end{figure}

Figure \ref{fig: size-dependence} shows the size dependence of the ground-state and first-excitation energies in all subspaces, where strong quantum fluctuations exist contrary to the Ising limit.
We have confirmed that numerical overlap between the ground and first-excited states in the same subspace is negligiblely small, $\av{\braket{0|1}}\lesssim 10^{-5}$.
In all panels, we can see the convergence of energies to approximately two or three levels with increasing the system size.

To check the degeneracy more apparently, we present the ground-state and first-excitation energies of the $N=64$ AXYSC in \Figref{WeakMajorana ZM}.
In contrast to the point A, we can see four-fold degeneracy in the points B and C, whereas both the points A and C belong to the (light and dark) green regions where the ground states are degenerate between $\bm{\chi}_0$ and $\bm{\chi}_1$ subspaces.
Hence, the degeneracy is unchanged even in the presence of quantum fluctuation.

\begin{figure}[h]
	\centering
	\includegraphics[width=\linewidth]{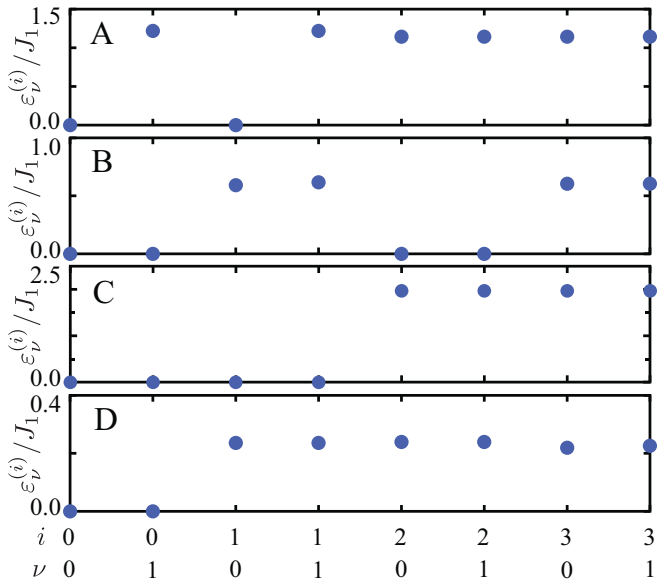}
	\caption{Ground-state and first-excitation energies obtained by the VMPS method in the $N=64$ AXYSC with the PBC. Panels from top to bottom correspond to the energy spectra at the points A, B, C, and D in \Figref{phase-diagram}. The horizontal axis represents index of subspace ($i=0,1,2,3$) and state number $\nu=0, 1$; i.e., ground state ($\nu=0$) and first excitation ($\nu=1$) in the $\bm{\chi}_i$ subspace.}
	\label{fig: WeakMajorana ZM}
\end{figure}

In the dark green region, as we explained in Sec.~III, there is a weak Majorana ZM in addition to the strong Majorana ZM in the Ising limit, indicating that some excited states are only two-fold degenerate.
At least in a certain region near the Ising limit, the excited states are still two-fold degenerate, while degeneracy of the ground states should be four-fold even if quantum fluctuations exist.
Therefore, we conclude that the weak Majorana ZM survives not only in the classical Ising limit but with quantum fluctuations, although we cannot deny the existence of two strong Majorana ZMs, i.e., a shift of Majorana ZM from weak to strong mode, in a part of the dark green region.

On the other hand, we find ground-state degeneracy even in the non-topological phase without Majorana ZMs (orange region in \Figref{phase-diagram}).
This degeneracy in the same subspace does not exist in light green region, which corresponds to the topological phase in the Kitaev chain with only first-neighbor terms.
Therefore, this degeneracy is also induced by the second-neighbor interaction.
In fact, one of the orange regions appears near isotropic spin condition, $\lambda_1= 0$ ($\theta= 0$) and $\lambda_2= 0$, and the other seems to be continuously connected with another isotropic point with the condition that $J_1= 0$ ($\theta= \pi/2$) and $\lambda_2= 0$.
In the isotropic case with large enough second-neighbor interaction, the ground states are doubly-degenerate dimerized states with spontaneous breaking of translational symmetry similar to the Majumdar--Ghosh state~\cite{Majumdar1969,Majumdar1970,Hikihara2001}.
This degeneracy in the same subspace can be characterized as a (at least weak) trivial ZM.
In fact, the origin of this trivial ZM, which is sublattice degree of freedom in doubly-extended unit cell with spontaneous breaking of translational symmetry induced by the second neighbor interactions, is common to the weak trivial ZM in the dark green and blue regions (see also Sec.~III).
Thus, considering the generality in physics, it may be appropriate that the weak ZM is regarded as not a Majorana ZM but just a trivial ZM.
However, it is more intriguing that if there is at least one Majorana ZM, other trivial ZMs can also play the role of Majorana ZM (see Sec.~III and \Appref{Majorana-pair2}).

Finally, we briefly mention effects of the boundary conditions.
In the present study, we determine the topological phase diagram with OBC to reduce numerical costs, while the degeneracy is investigated with PBC.
As explained in Sec.~III, the strong Majorana ZM and the corresponding degeneracy appear under both OBC and PBC in the AXYSC contrary to the Kitaev chain.
This significant feature originates from the non-local Jordan--Wigner transformation, and thus the strong Majorana ZM is a characteristic of bulk in the AXYSC.
Hence, with OBC, we can obtain the topological phase diagram regarding the strong Majorana ZM.
On the other hand, the weak Majorana ZM and the degeneracy emerge under only PBC at least in the Ising limit.
Therefore, to confirm the weak Majorana ZM, we have to examine whether additional degeneracy does appear or not with PBC.
In the thermodynamical limit, the bulk state is common between OBC and PBC, so that the phase boundaries determined using the strong Majorana ZM as a bulk property should be the same between OBC and PBC.
In other words, even with OBC, there is a certain topological invariance in the bulk, such as a winding number, which is difference between the light and dark green regions, while the weak Majorana ZM as an edge mode vanishes.


\section{Summary}

We have investigated Majorana ZMs in the AXYSC with the second-neighbor interaction.
In the Ising limit with PBC, we first analytically demonstrate two types of Majorana ZMs, i.e., strong and weak Majorana ZMs.
The weak Majorana ZM is restricted into a Hilbert subspace including the ground states, whereas the strong Majorana ZM appears in the whole Hilbert space.
With quantum fluctuations, we have found several topological phases characterized by the strong Majorana ZM by means of the VMPS method, in addition to a non-topological phase.
Furthermore, based on analysis of lowest excitations in each subspace, we have confirmed a weak Majorana ZM appearing with the strong Majorana ZM in topological phases induced by a large second-neighbor interaction.
Although multiple Majorana ZMs have already been reported, coexistence of mixed types of Majorana ZMs induced by long-ranged interaction with strong correlations is of interest from theoretical point of view.
Based on the presence of the two-fold degeneracy in the non-topological phase, we conclude that it is inappropriate to regard the weak trivial ZM as a Majorana ZM even in a topological phase.
Nonetheless, we also insist that the weak trivial ZM can play the role of weak Majorana ZM if there is other Majorana ZMs.

Our study will contribute to realizing the Majorana ZM if quantum spin materials described by the present AXYSC model are synthesized.
Additionally, the weak Majorana ZM appearing in the ground state should induce intriguing low-temperature physics.
For instance, if there is a large gap upon the ground state, the effects of excited states are effectively ignored at low enough temperature as compared with the energy gap.
Therefore, temperature dependence can be a possible probe to detect an interesting feature due to the weak Majorana ZM, which behaves in the same manner with the strong Majorana ZM only at low temperature.
Furthermore, $S=1/2$ spin has the same quantum statistics with a hard-core boson.
Since the hard-core boson is experimentally investigated in optical lattices, a promising quantum system to detect the weak Majorana ZM can also be realized in the optical lattices.

\begin{acknowledgments}
We would like to thank H. Katsura for valuable discussions.
Numerical computation in this work was carried out on the Supercomputer Center at Institute for Solid State Physics, University of Tokyo and the supercomputers at JAEA.
\end{acknowledgments}


\appendix
\renewcommand{\thetable}{A\Roman{table}}

\section{MPO representation of Hamiltonian}
\label{app: MPO representation}
Although the MPO representation of Hamiltonian is used in the VMPS method, it is neither unique nor well known how to obtain.  
In this section, we explicitly show the MPO representation of Hamiltonians, $\ham_n^{\iB}$ and $\ham_F$ with OBC. 
Thanks to the detailed review~\cite{Schollwock2011}, the term of AXYSC $\ham_n^{\iB}$ is easily transformed into the MPO form, given by
\begin{equation}
\sum_{n=1,2}\ham_n^{\iB}=\bm{H}_1\mbf{H}_2\cdots\mbf{H}_{N-1}\bm{H}_{N}^T
\end{equation}
where the local matrix (vector) operators are given by,
\begin{align}
\bm{H}_1&=\begin{pmatrix}
0, & \bm{M}_1, & \bm{P}_1, & 1 
\end{pmatrix},\\
\mbf{H}_j&=\begin{pmatrix}
1 & & & \\
\bm{p}_j^T & \mbf{L} & & \\
\bm{m}_j^T & & \mbf{L} & \\
0 & \bm{M}_j & \bm{P}_j & 1 
\end{pmatrix},\\
\bm{H}_{N}&=\begin{pmatrix}
1, & \bm{p}_{2}, & \bm{m}_{2}, & 0
\end{pmatrix},
\end{align}
with the lower matrix 
\begin{equation}
\mbf{L}=
\begin{pmatrix}
0 & 0 \\
1 & 0 \\
\end{pmatrix}.
\end{equation}
Here, we define local vector operators ($j=1,2,\cdots,N$), 
\begin{align}
\bm{M}_{j}&=\frac{1}{2}
\begin{pmatrix}
J_1 S_{j}^- + \lam_1 S_j^+, & J_2 S_{j}^- + \lam_2 S_j^+ 
\end{pmatrix},\\
\bm{P}_{j}&=\frac{1}{2}
\begin{pmatrix}
J_1 S_{j}^+ + \lam_1 S_j^-, & J_2 S_{j}^+ + \lam_2 S_j^-
\end{pmatrix},
\end{align}
and
\begin{align}
 \bm{p}_{j}=
\begin{pmatrix}
 S_{j}^+, & 0 
\end{pmatrix},
 \bm{m}_{j}=
\begin{pmatrix}
 S_{j}^-, & 0 
\end{pmatrix}.
\end{align}
On the other hand, the auxiliary field to restrict the parity subspace is non-trivial.
Since the eigenvalue of $\pf^\alp$ is $\pm 1$, we pay attention to the following relation
\begin{equation}
\av{\pf^\alpha-\chi^\alpha} = 
\begin{cases}
-\pf^\alp +\chi^\alp & (\chi^\alp=+)\\
\pf^\alp -\chi^\alp & (\chi^\alp=-)
\end{cases}.
\end{equation} 
The auxiliary field is rewritten by 
\begin{equation}
 \ham_F=\mu\br{3-\sum_\alp\chi^\alp \pf^\alp}. 
\end{equation}
Thus, we can express the auxiliary field with the MPO representation as
\begin{equation}
  \ham_F=\bm{F}_1\mbf{F}_2\cdots\mbf{F}_{N-1}\bm{F}_{N}^T
\end{equation}
with
\begin{align}
\bm{F}_1&=\mu\begin{pmatrix}
-2\chi^xS_1^x, & -2\chi^yS_1^y, & -2\chi^zS_1^z, & 3
\end{pmatrix},\\
\mbf{F}_j&=\begin{pmatrix}
2S_j^x & & & \\
& 2S_j^y & & \\
& & 2S_j^z &  \\
& & & 1
\end{pmatrix},\\
\bm{F}_{N}&=\begin{pmatrix}
2S_N^x, & 2S_N^y, & 2S_N^z, & 1
\end{pmatrix}.
\end{align}
It is not necessary to obtain the MPO representation for another auxiliary field $\ham_L$, because the operator is just a (superposed) direct product of MPS.
For the first-excited state $\ket{1}$, the effect of $\ham_L$ reads
\begin{equation}
  \ham_L\ket{1}=\epsilon \braket{0|1}\ket{0}.
\end{equation}
Therefore, it is sufficient to calculate the inner product of two MPSs $\braket{0|1}$.

Although the MPO representation with PBC is more complicated than that with OBC, it is available via correspondence between a single chain with PBC and a double chain coupled with two edges [see \Figref{model}(b)].
In this mapping, the site index is alternately labeled in the first and second chains, taht is, in \Figref{model}(b),
\begin{align}
  &(\vS_1,\vS_2,\vS_3,\cdots,\vS_{N/2},\vS_{N/2+1},\cdots,\vS_{N-2},\vS_{N-1},\vS_{N}) \notag \\
  & \ \to (\vS_1,\vS_3,\vS_5,\cdots,\vS_{N-1},\vS_{N},\cdots,\vS_{6},\vS_{4},\vS_{2}).
\end{align}
Thus, in appearance, longer-ranged interactions up to fourth-neighbor site are required.
Note that in this form, the bond dimension of the MPO with PBC is approximately doubled as compared with OBC, and we also have to set the bond dimension of MPS much larger due to a large value of entanglement entropy.

\section{A pair of Majorana ZMs and number of them}
\label{app: Majorana-pair}
In general, supporsing that a certain parity operator $\mathcal{Q}$ obeys $[\mathcal{Q},\ham]=0$, $\{\mathcal{Q},\smz^a\}=0$, and $\mathcal{Q}^2=1$ for a Majorana ZM $\smz^a$, we can make a pair to the Majorana ZM by using unitary transformation,
\begin{equation}
\smz^b=\ex^{-\im\frac{\pi}{4}\mathcal{Q}}\smz^a\ex^{\im\frac{\pi}{4}\mathcal{Q}}=\im\smz^a\mathcal{Q},
\end{equation}
satisfying the Majorana condition $\{ \smz^\tau , \smz^{\tau^\prime} \} = \delta_{\tau,\tau^\prime}$ and commutation relation $[\smz^b,\ham]=0$.
In the Ising chain, this parity operator corresponds to $\mathcal{Q}=-2\im \gamma_N^b \gamma_1^a=-4S_N^xS_1^x\pf$ for $\smz^a=\gamma_1^a$ and $\smz^b=\gamma_N^b$.
Simultaneously, we can obtain another pair to the Majorana ZM, $(\smz^b)^\prime=\im\smz^a\pf\neq \smz^b$, provided that $\mathcal{Q}=\pf$, while this mode $(\smz^b)^\prime$ does not anticommute with $\smz^b$, i.e., these modes are essentially not independent.
Although the definition of the paired Majorana ZM is not unique, only one independent Majorana ZM can be constructed by one Majorana ZM as its pair.
Thus, the number of the independent Majorana ZMs $D_m$ is always even, and the degeneracy of eigenstates in the Hamiltonian corresponds to $2^{D_m/2}$.

\section{Transformation of trivial ZMs to Majorana ZMs}
\label{app: Majorana-pair2}
As mentioned in Sec.~III, we can rewrite several trivial ZMs to Majorana ZMs, if there is an intrinsic Majorana ZM $\smz_0^a$ obeying $\{\smz_0^a,\pf\}=[\smz_0^a,\ham]=[\pf,\ham]=0$ and $2(\smz_0^a)^2=1$.
Here, we explicitely present the transformation of the trivial ZMs to Majorana ZMs.
(In this section, we assume strong ZMs, though weak ZMs can also be considered with introducing a certain projection.)
Firstly, we define the $j$-th trivial ZM and its parity by $\zeta_j^a$ and $Z_j$ ($j=1,2,\cdots,D$) where $D$ denotes number of the trivial ZMs, satisfying $[\zeta_j^a,\smz^a]=[\zeta_j^a,\pf]=[\zeta_j^a,\zeta_{j^\pri}^a]=0$ (independence of trivial ZMs) and $\{\zeta_j^a,Z_j\}=[\zeta_j^a,\ham]=[Z_j,\ham]=0$ (zero energy excitations) with $(\zeta_j^a)^2=Z_j^2=1$ (Z$_2$ quantities).
The trivial ZM operator $\zeta_j^a$ flips an eigenvalue of the parity operator $Z_j$, so that these can correspond to Pauli matrices $\sigma_j^x$ and $\sigma_j^z$ acting on a Hilbert space consisting of independent pseudo spins.
Hence, there is another Pauli matrix $\sigma_j^y=\im \sigma_j^x \sigma_j^z$, which is regarded as a pair to the trivial ZM $\zeta_j^b=\im \zeta_j^a Z_j$ satisfying $\{\zeta_j^b,Z_j\}=[\zeta_j^b,\ham]=0$ and $\{\zeta_j^\tau,\zeta_{j^\pri}^{\tau^\pri}\}=2\delta_{j,j^\pri}\delta_{\tau,\tau^\pri}$ (see Table~AI).
These trivial ZMs can be transformed to Majorana ZMs with introducing a Jordan--Wigner phase $\prod_{k<j} Z_k$ and a flip of fermionic parity $\im\smz_0^a\pf$. i.e., $\smz_j^\tau=\im \smz_0^a\pf \zeta_j^\tau \prod_{k>j} Z_k /\sqrt{2}$ for $\tau=a,b$ and $j=1,2,\cdots,D$.
In addition, we have to add a factor $\prod_{j} Z_j$ into a pair to the intrinsic Majorana ZM $\smz_0^b=\im \smz_0^a \pf$ to anticommute with other Majorana ZMs.
Table~AI shows the correspondence of them, where every Majorana ZM obtained by the transformation is independent from the others, that is, $\{\smz_j^\tau,\pf\}=[\smz_j^\tau,\ham]=0$ and $\{\smz_j^\tau,\smz_{j^\pri}^{\tau^\pri}\}=\delta_{j,j^\pri}\delta_{\tau,\tau^\pri}$.
Consequently, we can obtain $(D+1)$ Majorana ZMs, if there is an intrinsic Majorana ZM and other trivial ZMs associated with any symmetry breaking.

\begin{table*}[htb]
\caption{Transformation to Majorana ZMs. The Majorana ZMs satisfy $\{\smz_j^\tau,\pf\}=[\smz_j^\tau,\ham]=0$ and $\{\smz_j^\tau,\smz_{j^\pri}^{\tau^\pri}\}=\delta_{j,j^\pri}\delta_{\tau,\tau^\pri}$. These are similar to the Jordan--Wigner transformation.}
  \begin{tabular}{c|| c | c | c}
    Pauli matrix & Majorana ZM & Trivial ZMs & Transformations \\ \hline\hline
    $\sigma^x$ (flip) & $\smz_0^a$ & $\zeta_j^a$ & $\smz_j^a=\im \smz_0^a\pf \zeta_j^a \prod_{k>j} Z_k /\sqrt{2}$ \\
    $\sigma^z$ (parity) & $\pf$ & $Z_j$ &  \\
    $\sigma^y$ (pair) & $\smz_0^b=\im \smz_0^a \pf \prod_j Z_j$ & $\zeta_j^b=\im \zeta_j^a Z_j$ & $\smz_j^b=-\smz_0^a\pf \zeta_j^a Z_j \prod_{k>j} Z_k /\sqrt{2}$ \\
  \end{tabular}
\end{table*}

\bibliography{refs-a}


\end{document}